# Pneumonia App: a mobile application for efficient pediatric pneumonia diagnosis using explainable convolutional neural networks (CNN)

Jiaming Deng[1,2,3,4#], Zhenglin Chen[1,2,3,4#], Minjiang Chen[1], Lulu Xu[2,3,4], Jiaqi Yang[5], Zhendong Luo[6]*, Peiwu Qin[1,2,3,4]*

1 Zhejiang Key Laboratory of Imaging and Interventional Medicine, Zhejiang Engineering Research Center of Interventional Medicine Engineering and Biotechnology, The Fifth Affiliated Hospital of Wenzhou Medical University, Lishui 323000, China

2 Tsinghua-Berkeley Shenzhen Institute, Tsinghua Shenzhen International Graduate School, Tsinghua University, Shenzhen, China

3 Institute of Biopharmaceutical and Health Engineering, Shenzhen International Graduate School, Tsinghua University, Shenzhen, Guangdong, China

4 Key Lab for Industrial Biocatalysis, Ministry of Education, Department of Chemical Engineering, Tsinghua University, Beijing 100084, China

5 State Key Laboratory of Urban Water Resources and Environment, School of Civil & Environmental Engineering, Harbin Institute of Technology (Shenzhen), Shenzhen 518055, China

6 Department of Radiology, The University of Hong Kong - Shenzhen Hospital, Shenzhen 518055, China

#These authors contributed equally to this work.

*Corresponding authors. E-mail: lzhend@163.com (D. Luo); pwqin@sz.tsinghua.edu.cn (P. Qin)

Abstract

Mycoplasma pneumoniae pneumonia (MPP) poses significant diagnostic challenges in pediatric healthcare, especially in regions like China where it's prevalent. We introduce PneumoniaAPP, a mobile application leveraging deep learning techniques for rapid MPP detection. Our approach capitalizes on convolutional neural networks (CNNs) trained on a comprehensive dataset comprising 3345 chest X-ray (CXR) images, which includes 833 CXR images revealing MPP and additionally augmented with samples from a public dataset. The CNN model achieved an accuracy of 88.20% and an AUROC of 0.9218 across all classes, with a specific accuracy of 97.64% for the mycoplasma class, as demonstrated on the testing dataset. Furthermore, we integrated explainability techniques into PneumoniaAPP to aid respiratory physicians in lung opacity localization. Our contribution extends beyond existing research by targeting pediatric MPP, emphasizing the age group of 0-12 years, and prioritizing deployment on mobile devices. This work signifies a significant advancement in pediatric pneumonia diagnosis, offering a reliable and accessible tool to alleviate diagnostic burdens in healthcare settings.

1. Introduction

Mycoplasma pneumoniae pneumonia (MPP) is a community-acquired infection that primarily affects children and young adults. It is characterized by pulmonary inflammation, caused by mycoplasma pneumoniae (MP) infection, and can involve bronchi, bronchioles, alveoli, interstitium, and other lung sites. Previous studies[1–15] have shown that MPP is the most common form of community-acquired pneumonia (CAP) in children aged 5 years and older in China, accounting for 10-40% of CAP cases in school-aged children and adolescents. A significant increase in MP infections in China since last September has led to persistent high pressure in pediatric, respiratory, and general internal medicine clinics across the country, exacerbated by the unique symptoms and diagnostic features of MPP[16].

The clinical diagnosis of MPP requires a combination of laboratory microbiology, serological testing, and imaging[17]. MP culture, MP nucleic acid test, and MP antibody test are laboratory microbiological and serological tests for MPP diagnosis, among which MP culture is regarded as the "gold standard"[18]. However, due to special condition requirements and the slow growth of MP, MP culture is unrealistic for clinical diagnosis. In addition, although the MP detection kit can assist in rapid qualitative diagnosis, its sensitivity and specificity are low, and the positive rate is only about 60%. Sometimes, testing for antibodies can take at least five days or even a week to detect[19].

The clinical manifestations of MPP usually include fever and cough, which may accompany symptoms such as headache, runny nose, sore throat, and earache. Fever is usually moderate to high, and persistent fever may indicate a more serious illness. The cough is usually severe and may resemble whooping cough, while some children, especially infants and young children, may wheeze. The first signs of lung involvement may not be obvious, but as the disease progresses, distinct symptoms such as breathing sounds and dry or wet rales may appear. Although imaging results are critical for assessing clinical severity and predicting prognosis, inexperienced physicians can misinterpret them, leading to misdiagnoses. In addition, interpreting chest X-ray (CXR) images is time-consuming and laborious. Given the effectiveness of artificial intelligence techniques, especially deep convolutional neural networks (CNNs), in detecting and diagnosing various types of pneumonia[20–28], it is natural to seek to utilize deep learning techniques to improve the efficiency and reliability of MPP diagnosis. As we have mentioned the significance of CXR images for pneumonia detection, detect the disease from CXR images. Additionally, implement our algorithms to create automated applications to help support radiologists. For this endeavor, we have developed a mobile app, named PneumoniaAPP, to achieve rapid, efficient, and accessible pediatric pneumonia detection using deep learning-based algorithms.

Deep learning refers to a class of machine learning algorithms inspired by the structure and function of the neurons. It utilizes artificial neural networks with multiple layers to learn and extract intricate patterns and representations from complex data. By automatically learning hierarchical features, deep learning models can achieve remarkable performance in tasks such as image recognition, natural language processing, and medical diagnosis. Recent advancements in CXR image analysis have predominantly followed the technical pathway of convolutional neural networks in the domain of deep learning [29–31]. Convolutional neural networks (CNNs) are a class of deep learning models specifically designed for analyzing visual data, such as images. They utilize convolutional layers to automatically extract relevant features from the input data, followed by pooling layers to downsample the extracted features. This hierarchical approach allows CNNs to learn complex spatial patterns and achieve state-of-the-art performance in various computer vision tasks. Our approach follows the state-of-the-art design principles of CNN. Furthermore, we leverage deep learning explainability techniques to utilize the discriminative features learned by the CNN model for accurately localizing and describing pneumonia lesions.

To train the CNN model, we gathered CXR images of children aged from 0-12 and diagnosed with MPP from multiple medical institutions across China. We combine our dataset (833 Mycoplasma-CXR images and 169 normal-CXR images) with a selected subset (669 normal, 858 bacteria, and 816 virus CXR images) of CXR images from a publicly available pediatric pneumonia dataset[32], which includes two other types of pediatric pneumonia: viral pneumonia and bacterial pneumonia. Subsequently, we partitioned the dataset into three distinct subsets, namely training, validation, and testing, maintaining an 8:1:1 ratio, to form the model's training and evaluation cohorts. Based on the Pneumonia dataset we have collected, we developed a CNN model aimed at predicting the likelihood of pediatric MPP and generating annotations of suspected lesion areas of pneumonia based on input CXR images. The best-performing model achieved an accuracy of 88.20%, an AUC of 0.9218, and an F1 score of 0.8824 on the testing dataset. To further enhance accessibility for respiratory physicians, we integrated our model along with the explaining visualization algorithm into the PneumoniaApp. The subsequent sections of this paper are structured as follows: In Section 2, we present our review of existing AI-assisted diagnostic approaches for multiple pneumonia types, across different age people and different types of pathogens. Section 3 outlines the theoretical underpinnings and elaborates on our methodology. Section 4 provides the comparison results of multiple state-of-the-art deep learning algorithms on the pediatric pneumonia dataset, along with the visualization outcomes of the model exhibiting superior performance, and showcases samples of the user interface from the PneumoniaApp demo.

2. Related Works

Recently, significant progress has been made in pneumonia detection using medical radiology, such as CXR and computed tomography (CT), through the application of deep learning-based artificial intelligence techniques. Here, we introduce the most relevant studies pretraining to pediatric MPP diagnosis and discuss the motivation behind our research. Stephen et al.[33] utilized a new, straightforward model to effectively perform optimal classification tasks using deep neural network architectures. The model, tailored for pneumonia image classification, relies on convolutional neural network algorithms to extract relative features from images through convolution with a set of neurons. Similarly, Sharma et al. [34] also proposed different CNN architectures to automatically extract features from CXR images and classify normal CXR images from pneumonia CXR images. Kundu et al. designed an ensemble model using weighted averaging of three classification networks, GoogLeNet, ResNet-18, and DenseNet-121, to classify pneumonia CXR images from normal images, adopting the Weighted averaging ensemble technique was adopted to achieve better performance. The shortcomings of this research work lie in the failure to detect potential differences in the radiographic appearances of pneumonia caused by different pathogens on CXR images, which are crucial for the clinical diagnosis of various types of pneumonia. Additionally, there are significant differences in discernible features between CXR images of adult patients and those of children, which may lead to biases in the models towards these two populations. As pneumonia is more deadly for children, there is a more urgent need for computer-aided diagnostic methods tailored to pediatric pneumonia to differentiate between different types of pneumonia cases and normal cases.

In these directions, several researchers conducted more specific studies on pneumonia in children or differentiated pneumonia pathogens. Liz et al. [35] utilized two different datasets, one being the X-ray Pediatric Pneumonia (XrPP) dataset provided by Ben-Gurion University, and the other being a publicly available dataset of pediatric CXR images. The datasets comprise 950 annotated CXR images, categorized for children aged 1 to 16 years, proposed for training models assisting in pediatric pneumonia diagnosis. On this dataset, six different CNN architectures were considered, and an ensemble model was built for diagnosing pediatric pneumonia. Arun et al. [36] utilized the publicly available pediatric pneumonia dataset from Kermany et al. [32], which includes CXR images of children aged 1 to 5 years. They preprocessed the collected CXR images, including resizing and normalization, and introduced geometric transformations (such as rotation, scaling, and flipping) as data augmentation to prevent overfitting. They employed pre-trained deep CNN architectures and added channel attention modules after the last convolutional block of these models. Finally, kernel principal component analysis was used to reduce the dimensionality of features extracted from the channel attention module, and a stacking classifier combining various machine learning algorithms, including support vector classifier (SVC), logistic regression, k-nearest neighbors (KNN), Nu-SVC, and XGBClassifier, was employed for the final classification task. These two research endeavors overlooked MPP, a prevalent form of pediatric pneumonia, and employed methods that were overly intricate, diverging from the state-of-the-art practices of contemporary deep learning techniques. Alternatively, for pneumonia attributed to MP infection, Serener et al. [37] proposed leveraging deep learning techniques and CNNs to discern discrepancies between MPP and viral pneumonia, such as COVID-19, in CT images. Similar to the aforementioned studies, the study compared the performance disparities among convolutional network architectures like ResNet-50 and DenseNet-121 on CT image datasets. CT scans boast higher sensitivity and detail resolution in clinical pneumonia diagnosis compared to CXR, facilitating clearer visualization of lung lesions. However, this advantage is counterbalanced by heightened costs and risks associated with patients' exposure to elevated radiation doses. Moreover, the availability and accessibility of CT scans may be constrained in resource-limited medical settings. Consequently, for pediatric pneumonia diagnosis, methods reliant on CT images often struggle to meet practical application requirements.

We noted that only a few of the aforementioned deep learning-based pneumonia diagnostic methods utilized interpretable deep learning techniques [38,39], and there seems to be an insufficient in-depth analysis of their results. This finding suggests that, although interpretability techniques are crucial for understanding model decision processes and can help better evaluate and optimize the performance and reliability of deep learning models in pneumonia clinical diagnosis, there is limited consideration in the current research literature for the specific challenges encountered in the real-world implementation and deployment of deep learning techniques. This constitutes the second motivation behind our research, namely, utilizing deep learning interpretability techniques to facilitate the localization of lesions in pediatric MPP.

Most studies relied on deep CNN architectures like GoogLeNet and DenseNet-121, known for their large parameter counts and computational demands, particularly on mobile devices. Additionally, the common use of ensemble learning techniques further complicates matters for mobile deployment. Our third motivation is to address these challenges and enhance the efficient deployment of deep learning algorithm-based mobile applications in real-world scenarios.

3. Materials and Methods

In this section, we present our approach to constructing the dataset, developing pediatric pneumonia prediction models, and implementing them on an Android app.

Dataset

3.1 Data collection

The pediatric pneumonia CXR image dataset used for constructing our dataset comes from multiple sources, including two hospitals in China and a publicly available dataset from the internet [36]. We obtained CXR images of patients aged 0 to 12 diagnosed with Mycoplasma pneumoniae pneumonia from the two hospitals. To train CNNs capable of diagnosing pediatric Mycoplasma pneumoniae pneumonia, we retrieved CXR image data of three types of pediatric pneumonia: normal cases, bacterial pneumonia, and viral pneumonia, from the aforementioned publicly available dataset. To ensure class balance in our dataset, we randomly sampled images from each category of the original dataset and added them to ours. Ultimately, our training dataset consists of four categories of samples: normal, bacterial, viral, and Mycoplasma. After merging and randomly shuffling, we split them into training, validation, and testing sets in proportions of 80%, 10%, and 10%, respectively. The distribution of the four categories in each dataset split is shown in Table 1.

|  | Normal | Bacterial | Virus | Mycoplasma | total |
|---|---|---|---|---|---|
| Training | 670 | 686 | 652 | 666 | 2674 |
| Validation | 83 | 85 | 81 | 83 | 332 |
| Testing | 85 | 87 | 83 | 84 | 339 |
| Total | 838(169) | 858 | 816 | 833 | 3345 |

Table 1. Category distribution in the constructed Pneumonia dataset. The number within parentheses for the NORMAL category represents the number of normal samples collected by us.

3.2 Data Preprocessing

To mitigate the interference of irrelevant factors such as acquisition devices and storage formats on the learning process, which may impede the neural network from effectively discerning features useful for distinguishing between different types of pneumonia in the image data, we performed a series of preprocessing steps on the image data. Firstly, to standardize the interpretation of pixel values across images stored with different bit depths, we scaled the pixel values in each image to the range of 0 to 1 using the scaling formula:

$$x'_{ij} = \frac{x_{ij} - \min(X)}{\max(X) - \min(X)}$$

Here, $x_{ij}$ and $x'_{ij}$ represents the pixel value at position ij in the pixel matrix, and $X$ is a collection of the pixel values in the image. Next, we resized all images to

224x224 dimensions using bilinear interpolation and duplicated the grayscale images twice along the channel dimension to meet the requirements of the model input. Subsequently, we applied contrast enhancement to the image data to highlight details using the Contrast Limited Adaptive Histogram Equalization (CLAHE) technique. CLAHE is particularly effective for X-ray images because they often utilize continuous exposure, where low-level exposure is administered until the region of interest is identified[40]. Compared to more advanced dynamic contrast enhancement algorithms [41], CLAHE's superior real-time performance is better suited for our application scenario, which involves deployment in a mobile app. Finally, to ensure that the pixel values in the dataset follow the same distribution, we standardized the data across the three channels using mean and standard deviation according to:

$$x'_{ijc} = \frac{x_{ijc} - \mu_c}{\sigma_c}$$

Where $i, j$ has the same meaning as in equation (a), $c$ represents the channel dimension, $\mu_c$ denotes the mean of the dataset on the channel, and $\sigma_c$ represents the standard deviation of the dataset on the channel.

3.3 Data augmentation

Data augmentation is the process of expanding the dataset by applying random, non-degrading transformations to the images. Integrating data augmentation strategies into the data processing pipeline can reduce the risk of overfitting the model due to the limitation of dataset size. We implement automatic data augmentation on the training dataset using torchvision API. The selection of augmentation transformations is largely influenced by considerations of perturbations caused by the mobile phone camera on CXR images. Notably, during validation and testing, non-random data augmentation transformations will be used instead of random ones to ensure fair performance evaluation across different models. Figure 1. illustrates the results of applying the specified data augmentation transformations on pediatric pneumonia CXR images.

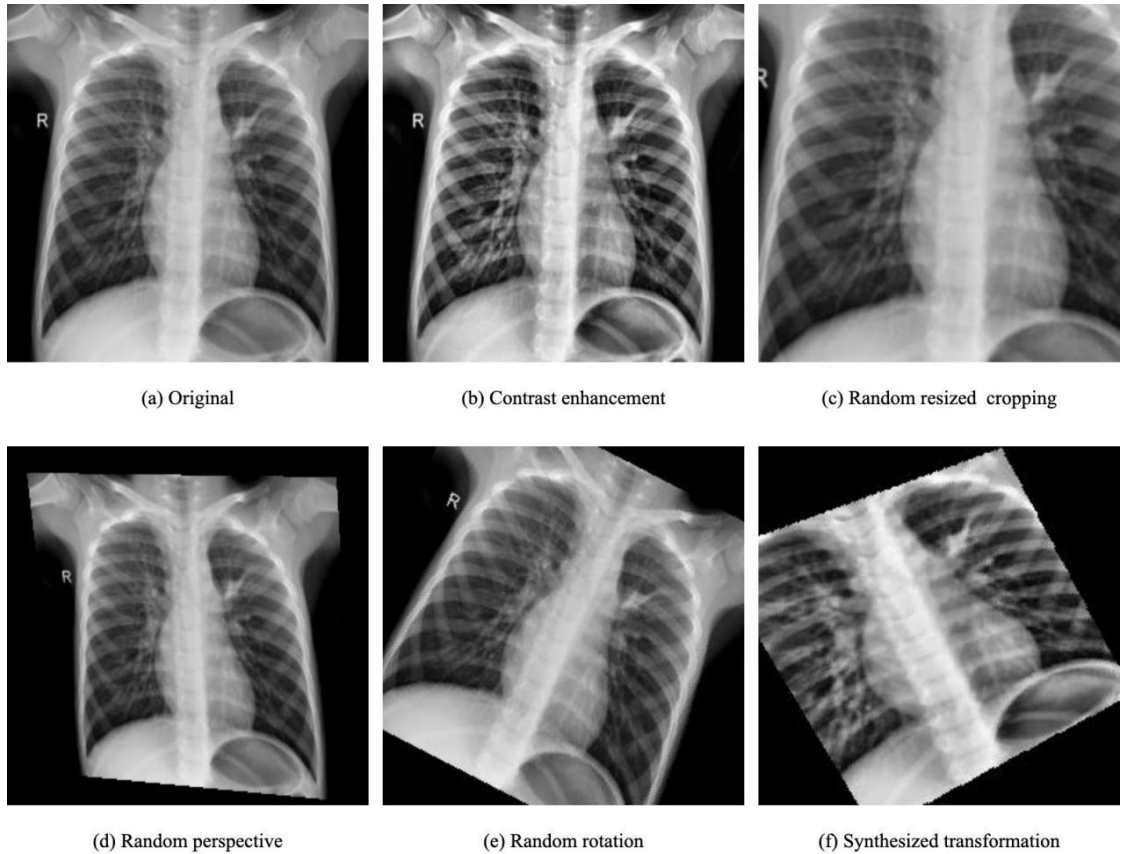

Figure 1. Example of original and augmented CXR images

3.4 Algorithm Architecture

We employ convolutional neural networks (CNN) to train CXR image classifiers for pediatric pneumonia prediction. A CNN [42] is a type of artificial feedforward neural network designed to extract features from data using convolutional structures. CNNs are specifically designed to automatically extract features from images, eliminating the need for manual feature extraction, unlike traditional methods. They are extensively employed in AI-assisted diagnosis tasks such as medical image classification, object detection, 3D reconstruction, and more.

One of the representatives of modern deep convolutional networks is ResNet. ResNet introduces residual connections in the network architecture, addressing the vanishing gradient problem during the training of deep neural networks, thereby facilitating the training of deeper networks. After demonstrating robust feature extraction performance on the ImageNet natural image dataset, ResNet has been widely used for automatic feature extraction in various tasks. In Section 4.2, we employed a lightweight variant of ResNet, ResNet-18, to train deep-learning models on the pediatric pneumonia CXR dataset. ResNet-18 consists of 8 basic blocks plus input and output layers, where each basic block comprises two convolutional modules with slight variations. Figure 2. illustrates the network architecture of ResNet-18.

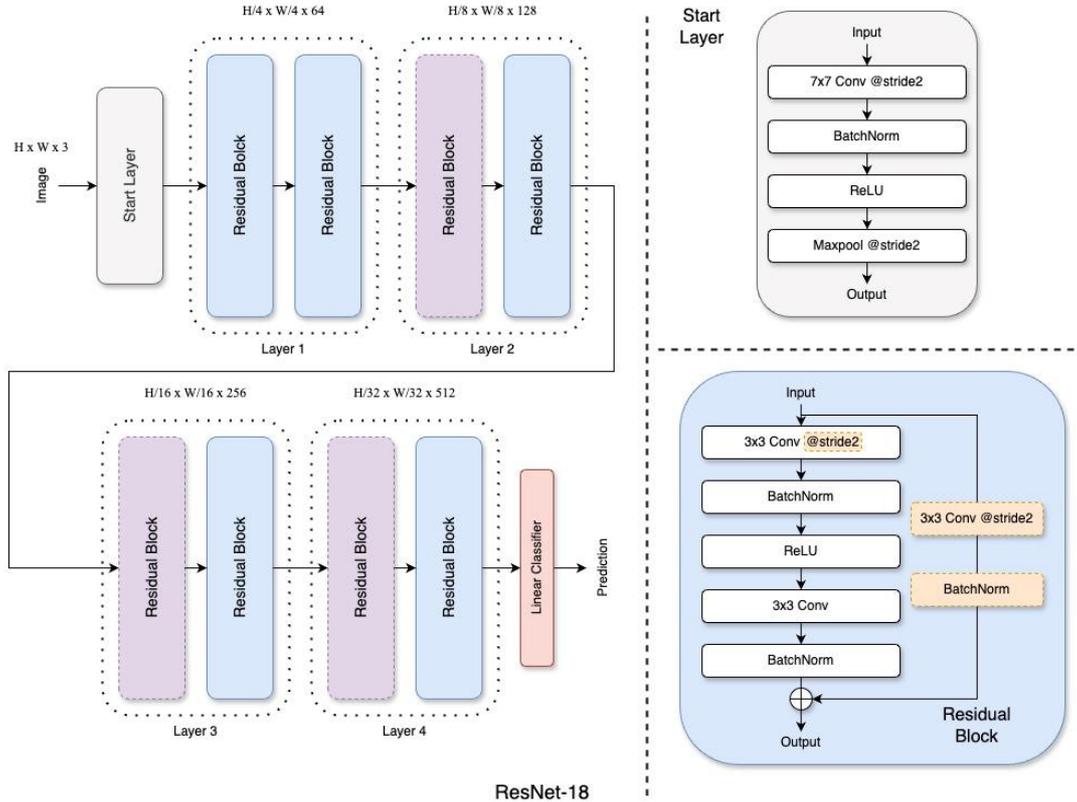

Figure 2. Architecture of ResNet-18. Layer 1 comprises two consecutive regular Residual blocks, followed by three subsequent layers, each containing a Residual block that downsamples the feature maps to one-fourth of the original size, alongside another regular Residual block. The Residual block illustration highlights modules and operations within dashed boxes, exclusive to downsampling Residual blocks, also denoted by purple dashed boxes in the overall ResNet-18 pipeline.

To further investigate the performance differences in feature extraction of classification network backbones on pediatric pneumonia CXR images, we have selected additional lightweight variants of network architectures based on the findings from the study [43]. In addition to ResNet-18, we have included EfficientNet (Tan & Le, 2019), RegNet[44], ConvNeXt[45], and Swin Transformer[46] as alternative network architectures. Considering the limitation for model size and computational burden when deploying in a mobile app, we comprehensively compare the parameters, computational complexity, and prediction performance metrics of the five models obtained.

Among them, Swin Transformer is the representative of vision transformers. Since ViT[47,48], i.e. the vanilla vision Transformer, introduced Transformer [49] into vision recognition tasks, various Transformer architectures were invented and achieved state-of-the-art performance on such tasks, and therefore become extremely popular. Here we selected a light-weighted variant, SwinV2-Tiny, to train a pediatric pneumonia detection model. The architecture of SwinTransformer-Tiny is presented in Figure 3.

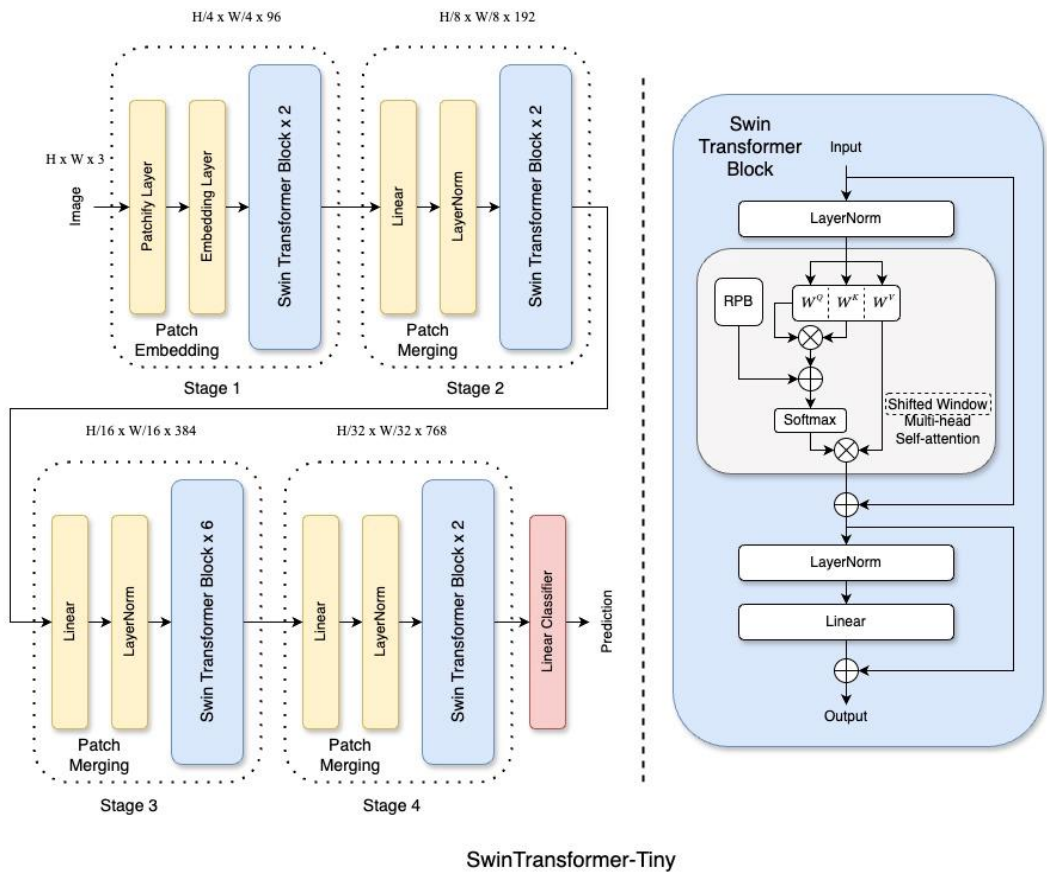

Figure 3. Architecture of SwinTransformer-Tiny. In Stage 1, the image is patchified, followed by token embedding through an embedding layer, and two consecutive Swin Transformer blocks. The first block uses standard multi-head self-attention, while the second employs shifted window multi-head self-attention. In subsequent stages, feature maps are merged (with linear layer and layer normalization module) before entering multiple Swin Transformer blocks for the next stage. Notably, Swin Transformer block sequences alternate between standard multi-head self-attention and shifted window versions in each stage.

ConvNeXt represents a modern CNN architecture. By incorporating modern CNN design elements such as the inverse bottleneck replacing activation functions with GELU, and adopting a stage ratio design similar to that of the Swin Transformer, ConvNeXt has demonstrated superior performance over visual Transformers in various visual recognition tasks. Similarly, we have chosen the lightweight variant of ConvNeXt, known as ConvNeXt-Tiny, for our study. The architecture diagram of ConvNeXt-Tiny is presented in Figure 4.

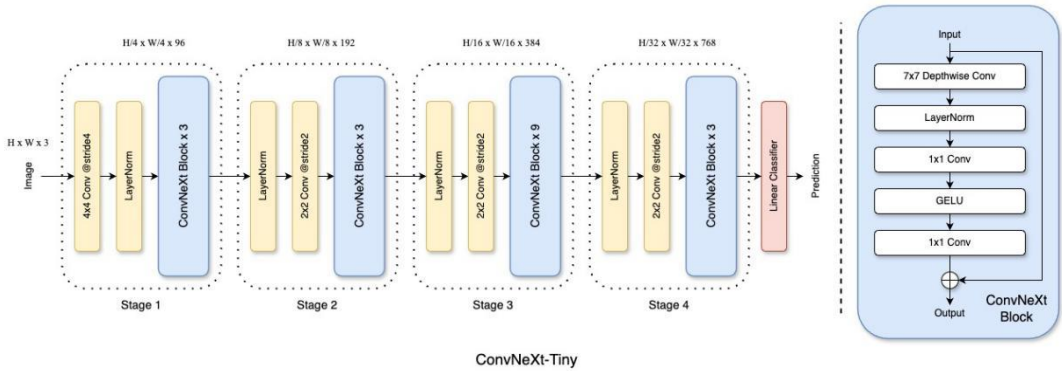

Figure 4. Architecture of ConvNeXt-Tiny. The pipeline comprises four stages. In Stage 1, the input image undergoes patchification via non-overlapping 4x4 convolution layers, followed by layer normalization. In the subsequent three stages, feature maps from Stage 1 undergo layer-wise normalization, followed by downsampling via 2x2 non-overlapping convolution layers with a stride of 2, and then enter a sequence of ConvNeXt blocks. Depthwise convolution is achieved using group convolution followed by 1x1 convolution layers. Further details of the experimental setup and results are presented in Section 4.2.

Class Activation Map (CAM) is a type of deep convolutional network feature visualization technique, which provides visual explanations for a single input by using a linear weighted combination of activation maps from convolutional layers. CAMs are commonly used to help explore the working mechanism and decision-making basis of deep convolutional neural networks.

CAMs can be formalized as:

$$L^c_{CAM} = ReLU(\sum_{k=1}^{N_l} \alpha_k A_k)$$

Where $c$ represents the target class, $A_k$ denotes the feature map, $\alpha_k$ represents the corresponding weight coefficient of the feature map, $N_l$ denotes the number of feature maps in the $k=1$ layer, and $ReLU$ is the commonly used activation function in CAM, defined as:

$$ReLU(x) = \max(0, x)$$

Typically, CAMs output matrices of the same size as the original image, with values distributed between 0 and 1. It is usually displayed as a grayscale image, where the value at each pixel position indicates the importance of the corresponding feature in the original image for network classification.

We utilize CAMs to provide indications of distinguishable feature regions in the original image, applied in the pediatric pneumonia prediction model. In Section 4.3, we present qualitative results of different CAM technique variants in the pediatric pneumonia prediction model. We observe that regions with higher heatmap intensity in CAM are primarily concentrated in the shadow areas of the lung in the original image, providing additional auxiliary information for respiratory specialists in the diagnosis of pediatric Mycoplasma pneumoniae pneumonia.

3.5 Transfer Learning

For training deep learning models, larger datasets typically imply better generalization performance. However, for the prediction of pediatric MPP, which is the focus of this study, the scarcity of data makes it impractical to increase the dataset size. To address this issue, we employ transfer learning techniques, making full use of model weights pre-trained on larger datasets such as ImageNet-1K. We initialize our models using the available weights from the pre-trained models. Consequently, models initialized with pre-trained weights converge faster during training and are less prone to overfitting.

3.6 Mobile Device Deployment

To facilitate the application of our pediatric pneumonia prediction model in real-world scenarios, we have developed an Android application prototype named "PneumoniaAPP." To deploy the pediatric pneumonia prediction model on mobile devices, we wrap it as a callable interface using the Python Flask framework and call these interfaces in the Android development environment. The user interface of PneumoniaApp is illustrated in Figure 5.

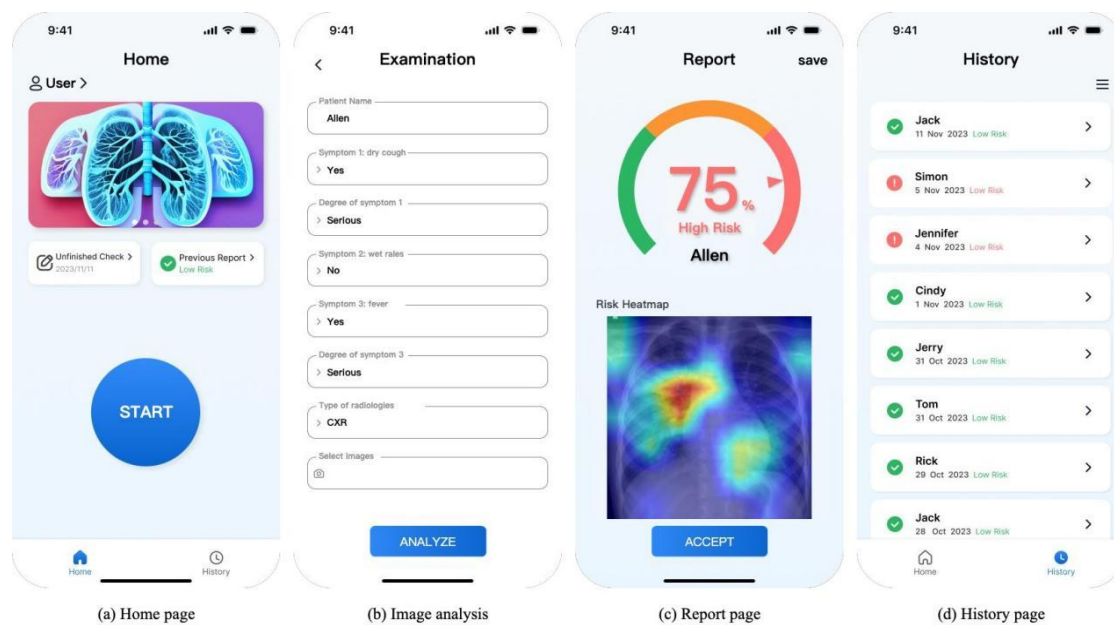

Figure 5. PneumoniaApp demo user interface example.

4. Results and Discussion

In this section, we investigate two main aspects: (1) the impact of different network architectures on the performance of pediatric pneumonia prediction models, evaluated through metrics including accuracy, recall, precision, AUC, and F1 score, and (2) the influence of various CAM algorithms on the visualization results of model outputs. Our experiments were conducted on a single RTX 2080Ti GPU in a Windows 10 environment.

4.1 Transfer Learning Performance

Given the constraints posed by the computational capabilities of mobile devices, our objective is to choose model architectures that efficiently extract image features while demanding a minimal volume of parameters and computational resources. Therefore, as discussed in Section 3.2, we leverage insights from the study [43]to comprehensively assess factors such as prediction accuracy on the ImageNet dataset, availability of pre-trained weights, transferability, model parameters, and computational burden on devices. Based on these criteria, we select five lightweight architectures: ResNet-18, RegNetX-400mf, EfficientNet-B0, SwinV2-Tiny, and ConvNeXt-Tiny. The experiments conducted in this subsection are intended to further compare the transfer learning performance of these five architectures on the pediatric pneumonia dataset.

For each architecture involved in the experiments, we employ backpropagation on the training dataset to update the model parameters and optimize the hyperparameters in the training settings based on their performance on the validation dataset. Following the construction of model architectures using PyTorch APIs, we initialize the parameters using pre-trained weights (version "ImageNet-1K-V1") provided by the Torchvision library. We replace the original architecture's fully connected layer, which outputs a shape of 1000x1, with another fully connected layer that is randomly initialized and outputs a 4x1 tensor for predicting the four categories of CXR images. We utilize the Adam optimizer to update the model weights, with the initial learning rate set to 1E-4 and the weight decay coefficient to 5E-3. Additionally, we reduce the learning rate by a factor of 0.1 every 50 epochs during training using a learning rate scheduler.

For image preprocessing and data augmentation procedure, we 1) firstly, apply CLAHE to raw CXR images, with clip_limit=2.0 and grid_size=(8, 8); 2) secondly, resize the contrast-enhanced image to Hx256, where H represents the height of the raw image, with its original height-weight ratio maintained; 3) thirdly, apply random cropping and resizing to the output image from the previous step, with the cropping ratio between 0.4 and 0.8 and resize it to a 224x224 resolution using bilinear interpolation; 4) for a probability of 0.6, apply a random perspective transformation to images from the previous step, where distortion scale is set to 0.4; 5) randomly rotate the images as a angle from -45 degrees to 45 degrees; 6) rescale the pixel value of the images to range [0, 1]; 7) lastly, normalize the pixel value of the image with a three channel mean value [0.485, 0.456, 0.406] of and std value of [0.229, 0.224, 0.225]. The pipeline of preprocessing and augmentation transformations applied to training image data is presented in Figure 6.

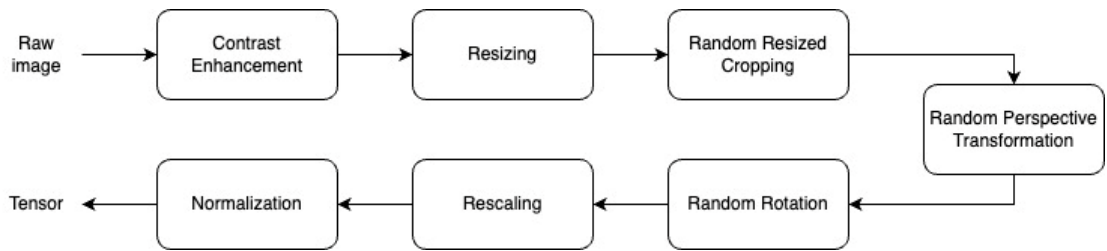

Figure 6. Preprocessing and augmentation pipeline for training data.

After multiple rounds of training, validation, and hyperparameter tuning, we obtained the optimal training hyperparameter settings described above. We then proceed to train the model with a maximum of 100 epochs, continuously saving and updating the model with the lowest validation set loss during the process. The model obtained after 100 epochs of training is considered the optimal model for the current architecture, and its performance is compared with models trained using other architectures. Figure 7. presents a validation loss line chart of the training process of these five models.

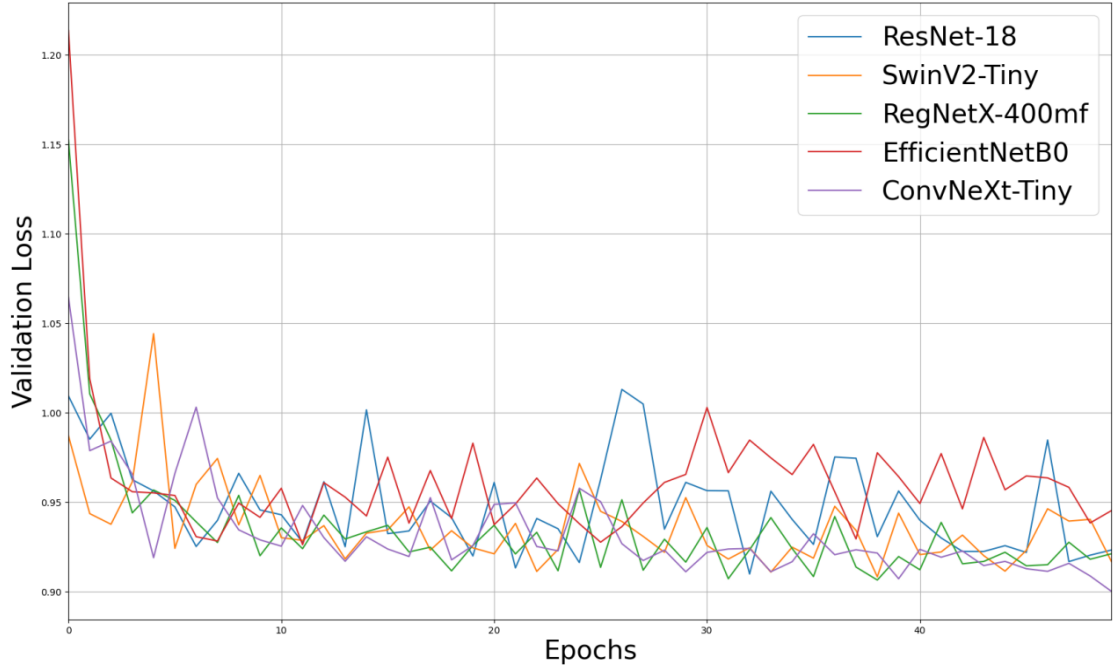

Figure 7. Evolution of the validation loss during training for ResNet-18, SwinV2-Tiny, RegNetX-400mf, EfficientNetB0, and ConvNeXt-Tiny model in first 50 epochs.

We evaluate the performance of our model on the test dataset using multiple indicative metrics, including accuracy, recall, precision, AUC, and F1 score, as indicators of its effectiveness. For binary classification tasks, where samples are classified into True positive (TP), false negative (FN), false positive (FP), and True negative (TN), the metrics are defined as follows:

$$recall = \frac{\text{TP}}{TP + FN}$$
$$precision = \frac{\text{TP}}{TP + FP}$$
$$accuracy = \frac{\text{TP} + \text{TN}}{TP + TN + FP + FN}$$
$$F1\ score = \frac{2 * \text{precision} * \text{recall}}{recall + precision} = \frac{2\text{TP}}{2TP + FP + FN}$$

To compute the Area Under the Curve (AUC), where P represents the set of all positive samples N represents the set of all negative samples, and rank_s represents the rank of sample s when all samples are sorted in ascending order according to the model's predicted score, AUC can be calculated as:

$$AUC = \frac{\sum_{s \in P} \text{rank}_s - \frac{|P|(|P|+1)}{2}}{|P||N|}$$

For the multi-class classification of pediatric pneumonia in our study, we utilize macro-averaged metrics, which represent the average of recall, precision, accuracy, AUC, and F1 score across all classes, providing a comprehensive measure of the model's performance.

The performance metrics results obtained on the test dataset for the optimal models of each architecture (Table) show that the ConvNeXt-Tiny model achieves a prediction accuracy of 88.20%, an AUC of 0.9218, and an F1 score of 0.8824, which is significantly superior to the other four architectures. Additionally, the ConvNeXt-Tiny model's parameter size and computational demand on devices are within acceptable limits. Therefore, we select the ConvNeXt-Tiny model for deployment in the PneumoniaApp

| Architecture | Acc. | Recall | Precision | Auc | F1 score | Number of Parameters | Computational cost (GFLOPs) |
|---|---|---|---|---|---|---|---|
| ResNet-18 | 82.01% | 0.8219 | 0.8385 | 0.8811 | 0.8201 | 11.7M | 1.81 |
| RegNetX-400mf | 82.89% | 0.8294 | 0.8424 | 0.8862 | 0.8335 | 5.5M | 0.41 |
| EfficientNet-B0 | 82.01% | 0.8205 | 0.8386 | 0.8803 | 0.8261 | 5.3M | 0.39 |
| SwinV2-Tiny | 84.07% | 0.8400 | 0.8473 | 0.8933 | 0.8400 | 28.4M | 5.94 |
| ConvNeXt-Tiny | **88.20%** | **0.8830** | **0.8844** | **0.9218** | **0.8824** | 28.6M | 4.46 |

Table 2. Performance of ResNet-18, RegNetX-400mf, EfficientNet-B0, SwinV2-Tiny, and ConvNeXt-Tiny optimal models. Boldface for best performer in the comparison.

4.2 Model Interpretability Study

We utilized the CAM techniques introduced in Section 3.2 to visualize the features extracted by the optimal models, aiming to pinpoint lesion regions. Here, we employed two popular CAM variants with distinct technical approaches: gradient-based Grad-CAM and score-based Score-CAM. For comparison, alongside the ConvNeXt-Tiny model, we also present the visualization features of the RegNetX-400mf model, which achieved impressive performance with an extremely low parameter size (5.5M) and minimal computational demands (0.41GFLOPS). Notably, as shown in the table, the RegNetX-400mf model achieved a prediction accuracy of 82.89%, an AUC of 0.8862, and an F1 score of 0.8335 on the test dataset.

We observed that the visualized feature maps (Figure 8. and Figure 9. ) of the two models effectively illustrate the rationale behind model classification, however, with distinct characteristics. The visualized feature maps of the RegNetX-400mf model obtained from both approaches are highly similar, displaying dispersed high-heat regions that almost cover the whole visible lung area in the CXR images. We speculate that the exceedingly low parameter size of the RegNetX-400mf model constrains its capacity to learn fine-grained features from the pediatric pneumonia CXR dataset.

In contrast, the visualized feature maps of the ConvNext-Tiny model exhibit distinct differences. Specifically, we observed that in the visualized feature maps generated using Grad-CAM, the high-heat regions are concentrated in areas lacking significant relevance (such as those of the Mycoplasma class). We posit that gradient-based CAM techniques may lead to a misinterpretation of the ConvNeXt model's features. Conversely, the visualized feature maps generated using Score-CAM effectively highlight suspected lesions corresponding to the pneumonia category. Based on these findings, we have integrated Score-CAM into the PneumoniaApp to provide doctors with reference information when interpreting pediatric CXR images and identifying lesion areas.

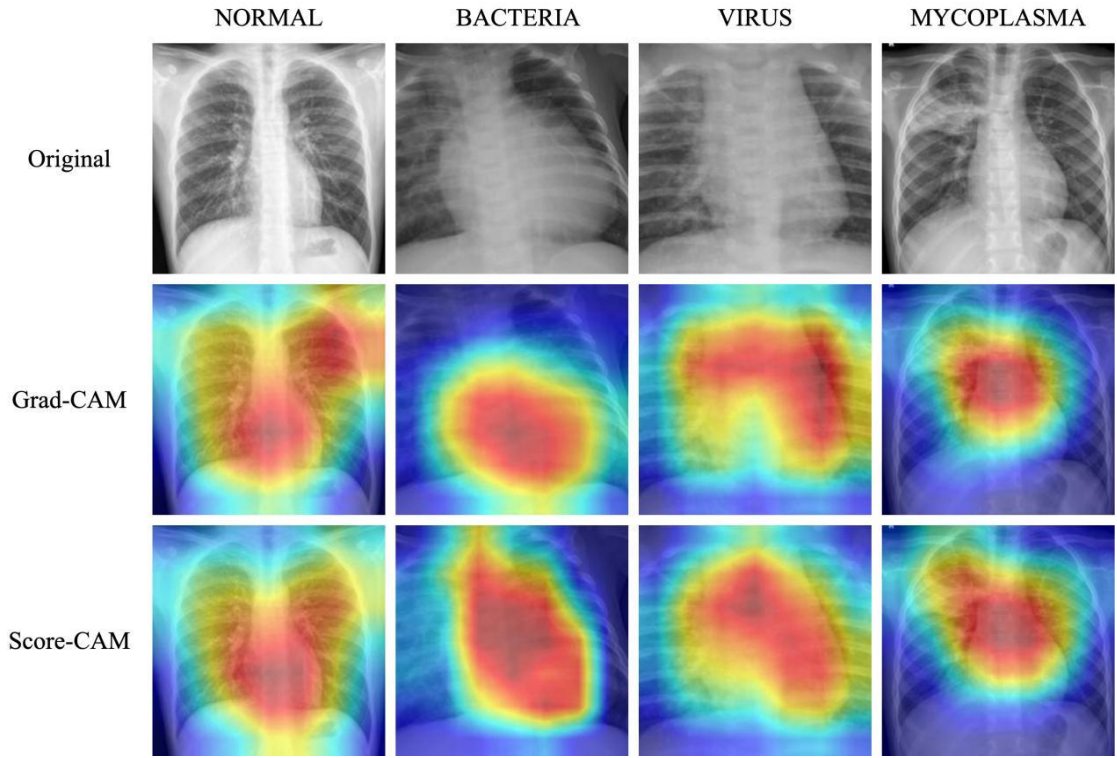

Figure 8. The feature visualization results of the RegNetX-400mf model.

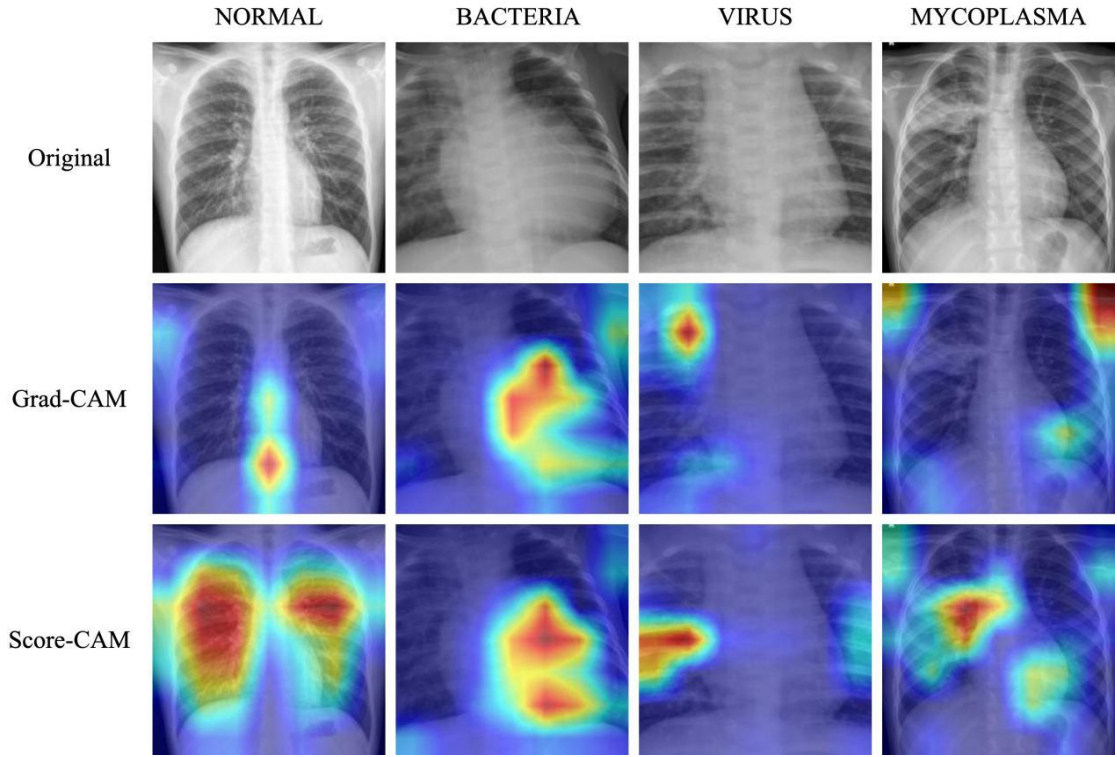

Figure 9. The feature visualization results of the ConvNeXt-Tiny model.

5. Conclusion

　　Pediatric Mycoplasma pneumoniae pneumonia is characterized by periodic outbreaks, imposing continuous diagnostic and therapeutic challenges on hospitals worldwide. In this study, we introduced an additional dataset specifically targeting MPP in children. We applied data augmentation techniques to pediatric pneumonia CXR image data to alleviate the risk of overfitting posed by limitations in dataset size. Leveraging advanced deep-learning visual algorithms, we developed a CNN model capable of identifying pediatric pneumonia types from CXR images. To ensure the model's efficiency and reliability, we compared the performance of several state-of-the-art lightweight convolutional network architectures on the proposed dataset, with the best performer achieving a prediction accuracy of 88.20% and an AUC of 0.9218. We employed CAM techniques to visualize the features learned by the pediatric pneumonia classification model, providing valuable guidance for respiratory physicians in lesion localization from CXR images. By deploying the aforementioned algorithms and models in a mobile application, we further facilitated their real-world application, greatly expediting the diagnosis of pediatric Mycoplasma pneumoniae pneumonia, reducing the risk of misdiagnosis, and thereby alleviating the diagnostic and therapeutic pressures on hospitals during specific periods.

　　Compared to previous research efforts using deep learning algorithms to assist in the diagnosis of pediatric pneumonia or mycoplasma pneumoniae pneumonia, our approach has several distinctive features: (1) We take into account the presence of mycoplasma, an important yet overlooked pathogen in pediatric pneumonia studies, thus reducing the likelihood of misdiagnosing mycoplasma pneumonia as pneumonia caused by other pathogens. (2) We focus on the age group of 0-12 years old, which is the most common and severely affected population in outpatient settings. (3) We conduct our research with the premise of deployment on mobile devices, ensuring the availability of the models and algorithms deployed in the Android application.

**Data and code availability**

All data and codes in this study are available from the corresponding authors upon request.

**Declaration of competing interest**

**Acknowledgments**

We thank the support from the National Natural Science Foundation of China31970752,32350410397; Science, Technology, Innovation Commission of ShenzhenMunicipality, JCYJ20220530143014032, JCYJ20230807113017035, WDZC20200820173710001; Shenzhen Science and Technology Program, JCYJ20230807113017035; Shenzhen Medical Research Funds, D2301002; Department of Chemical Engineering-iBHE special cooperation joint fund project, DCE-iBHE-2022-3; Tsinghua Shenzhen International Graduate School Crossdisciplinary Research and Innovation Fund Research Plan, JC2022009; and Bureau ofPlanning, Land and Resources of Shenzhen Municipality (2022) 207.